\documentclass[aip,preprint,12pt]{revtex4-1}
\usepackage{graphicx}
\usepackage{amsmath}
\usepackage{gensymb}
\begin{document}

\title{Measurement of the linear thermo-optical coefficient of Ga$_{0.51}$In$_{0.49}$P using photonic crystal nanocavities} 


\author{Sergei Sokolov}
\email[]{s.a.sokolov@uu.nl}
\homepage[]{http://www.nanolinx.nl/}
\affiliation{Nanophotonics, Debye Institute for Nanomaterials Science, Center for Extreme Matter and Emergent Phenomena, Utrecht University, P.O. Box 80.000, 3508 TA Utrecht, The Netherlands }
\affiliation{Complex Photonic Systems (COPS), MESA+ Institute for Nanotechnology, University of Twente, P.O. Box 217, 7500 AE Enschede, The Netherlands}
\author{Jin Lian}
\affiliation{Nanophotonics, Debye Institute for Nanomaterials Science, Center for Extreme Matter and Emergent Phenomena, Utrecht University, P.O. Box 80.000, 3508 TA Utrecht, The Netherlands }
\author{Sylvain Combri\'{e}}
\affiliation{Thales Research \& Technology, Route D\'{e}partementale 128, 91767 Palaiseau, France}
\author{Alfredo De Rossi}
\affiliation{Thales Research \& Technology, Route D\'{e}partementale 128, 91767 Palaiseau, France}
\author{Allard P. Mosk}
\affiliation{Nanophotonics, Debye Institute for Nanomaterials Science, Center for Extreme Matter and Emergent Phenomena, Utrecht University, P.O. Box 80.000, 3508 TA Utrecht, The Netherlands }

\begin{abstract}
\textbf{
Ga$_{0.51}$In$_{0.49}$P is a promising candidate for thermally tunable nanophotonic devices due to its low thermal conductivity. In this work we study its thermo-optical response. We obtain the linear thermo-optical coefficient $dn/dT=2.0\pm0.3\cdot 10^{-4}\,\rm{K}^{-1}$ by investigating the transmission properties of a single mode-gap photonic crystal nanocavity. 
}
\end{abstract}
\date{\today}
\maketitle 

\section{introduction}
Several III-V semiconductors are used for nanophotonic devices, such as GaAs \cite{Faraon2008}, GaP \cite{Rivoire2010}, InP \cite{Yu2013} for their specific properties such as optical tunability. Local tuning of the refractive index has been proposed as a method to create localized resonances in a waveguide \cite{Notomi2008optexpr}, and to tune complex localized states \cite{oxydation3, Pan2008}. For local thermal tuning, the high thermal conuctivity of Si or $\rm{Si_3N_4}$ is unfavorable. A potential material for thermally tunable nanophotonic devices is GaInP which became popular during recent years \cite{Sokolov2015, Lian2016, Martin2016, Combrie2009, Clark2013}. Its thermal conductivity \cite{adachi} is more than 6 times smaller than for Si and $\rm{Si_3N_4}$, which in addition to absence of two-photon absorption at 1550 nm and favorable nonlinear properties \cite{Combrie2009} makes it a promising candidate for thermally tunable photonics where multiple closely spaced elements have to be tuned independently. The thermo-optical coefficient of this material has indirectly been obtained before as a result of a series of complex measurements and heat diffusion calculations in Ref. \citenum{Sokolov2015}, which implies the need of a more direct measurement. To our knowledge the precise value of $dn/dT$ has not been reported in the literature. Therefore in this work we investigate the thermal response of a single photonic crystal nanocavity made of Ga$_{0.51}$In$_{0.49}$P for a homogeneously heated sample where temperature of the sample is directly measured. Our rigorous analysis allows us to obtain the precise value of the thermo-optical coefficient of refractive index for this material.

\section{Sample and experimental setup}
To experimentally measure the thermo-optical coefficient of the Ga$_{0.51}$In$_{0.49}$P a nanophotonic sample containing photonic crystal nanocavities was mounted on a thermally controlled stage, so the sample was homogeneously heated and the temperature of the stage was locked with a precision of $\pm0.001\,\rm{K}$. 

The temperature drop between the stage and the sample was measured in a separate run using a PT-100 temperature sensor placed in the position of the photonic crystal. It was found that temperature drop varied from 1.1 K to 1.4 K linearly for stage temperatures between 27.5 and 77.5 $\rm{^oC}$. All temperatures mentioned from here are sample temperatures corrected for this temperature drop and measured with an absolute precision of 0.3 K as verified with a calibrated digital thermometer.

\begin{figure}[htbp]
 \centering \includegraphics[width=8.4cm]{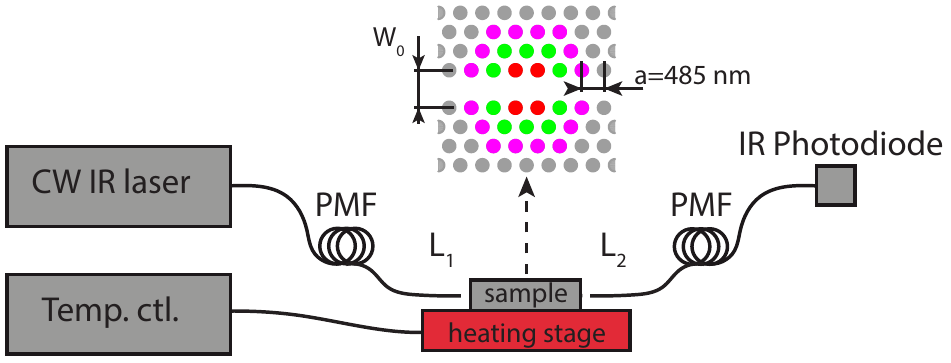}
 \caption{\label{fig1}\textbf{Experimental setup.} Transmission through the sample is measured. The sample temperature is locked to $\pm0.001\, \rm{^oC}$ using a temperature controller. Inset shows the cavity structure. Holes are represented with circles. Different colors correspond to different hole shifts. }
\end{figure}

The sample is an air-suspended photonic crystal membrane with a linear array of 10 directly coupled mode-gap photonic crystal nanocavities made of Ga$_{0.51}$In$_{0.49}$P. The thickness of semiconductor membrane is 180 nm. The structure of a single cavity is presented as the inset in Figure \ref{fig1}. It is made in a photonic crystal waveguide with width $W_0=0.98\sqrt{3}a$ where $a$=485 nm is a period of a triangular photonic crystal lattice. Red holes are shifted away from the waveguide by 6 nm, green - by 4 nm and purple ones are shifted by 2 nm to create a cavity mode. Such cavities are known for large experimentally measured Q-factors \cite{Kuramochi2006, Combrie2008} and therefore they are suitable for precise thermal measurements. The first and last cavities in the array are coupled to input and output photonic crystal waveguides with width $W_1=1.1\sqrt{3}a$. The waveguides are used to launch and collect light from the structure. 

\begin{figure}[htbp]
 \centering \includegraphics[width=8.4cm]{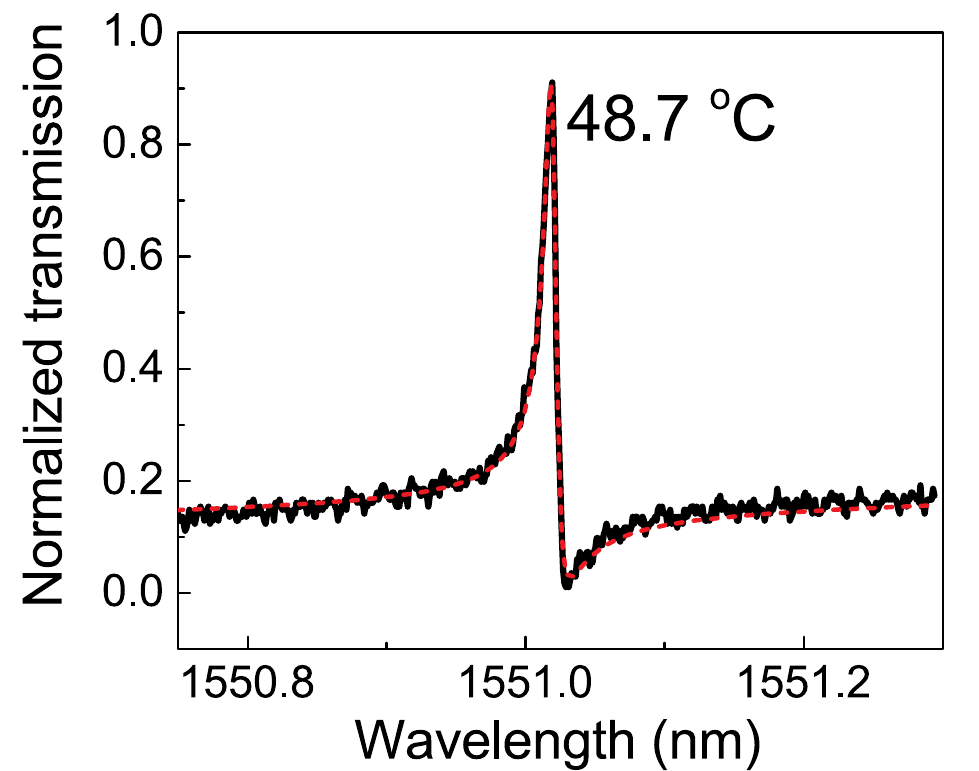}
 \caption{\label{fig2a}\textbf{Transmission spectrum.} Transmission spectra of the sample at 48.7~$\rm{^oC}$. Dashed line represents Fano lineshape fit.}
\end{figure}
The structure was probed by IR light with a wavelength around 1550 nm from the CW tunable laser. The light was coupled into and out-coupled from the structure with polarization-maintaining lensed fibers $\rm{L_1}$ and $\rm{L_2}$ with NA=0.55. The out-coupled light was collected on an IR photodiode for transmission measurements. The sample was kept in nitrogen atmosphere to avoid oxidation \cite{oxydation1,Navr1968}. The resonance frequencies of cavities are perturbed by unavoidable disorder, which breaks the resonance hybridization. This normally undesired effect allowed us to pick an isolated single cavity resonance. We picked the single resonance corresponding to the $\rm{3^{rd}}$ cavity in the array, which was verified by our pump line-scan technique \cite{Sokolov2015, Lian2016, Sokolov2016}. The transmission spectrum of the resonance at 48.7 $\rm{^oC}$ is presented in Figure \ref{fig2a}. The spectrum has a clear Fano-like lineshape due to the interference with transmitted TM light. The spectrum is fitted with a Fano lineshape function \cite{Fano1961, Zhou2014} and a $\rm{1^{st}}$ order polynomial for the background. The lineshape is perfectly described by the fit, so it is used to obtain line parameters. The loaded Q-factor of the resonance is $Q=1.6\pm0.1\cdot10^5$. 

\section{Experiment description}
\begin{figure}[htbp]
\centering \includegraphics[width=8.4cm]{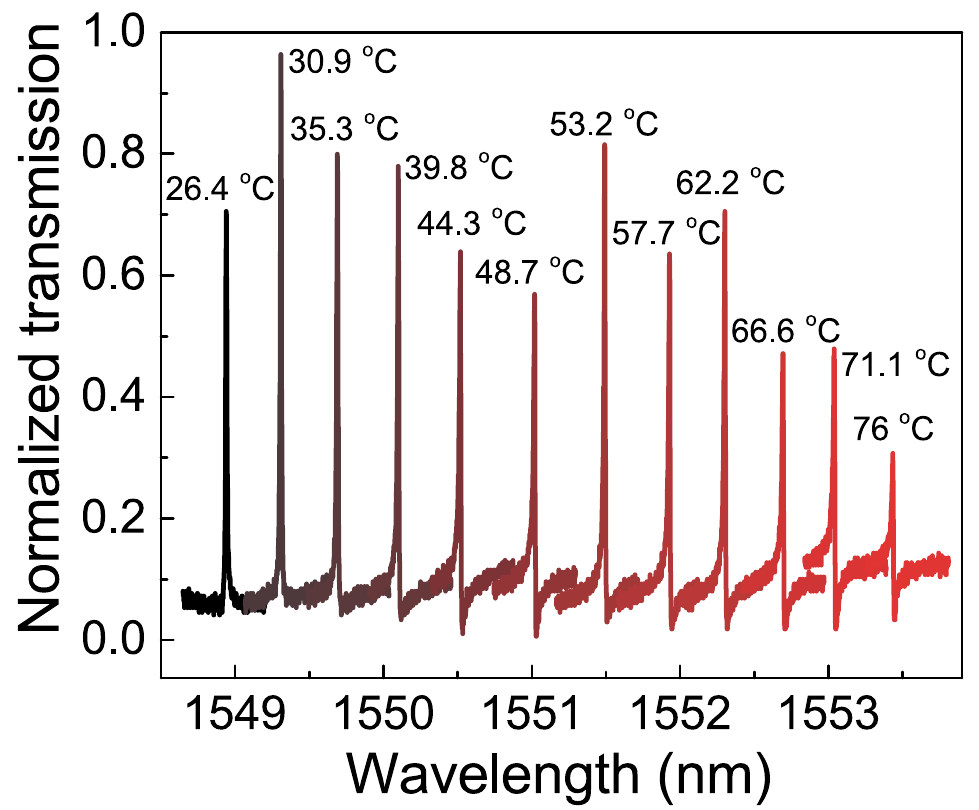}
\caption{\label{fig2}\textbf{Transmission spectra.}  Patched graph showing all transmission spectra collected in this experiment. Color of lines changes from black to red to emphasize the temperature difference.}
\end{figure}
\begin{figure}[htbp]
 \centering \includegraphics[width=8.4cm]{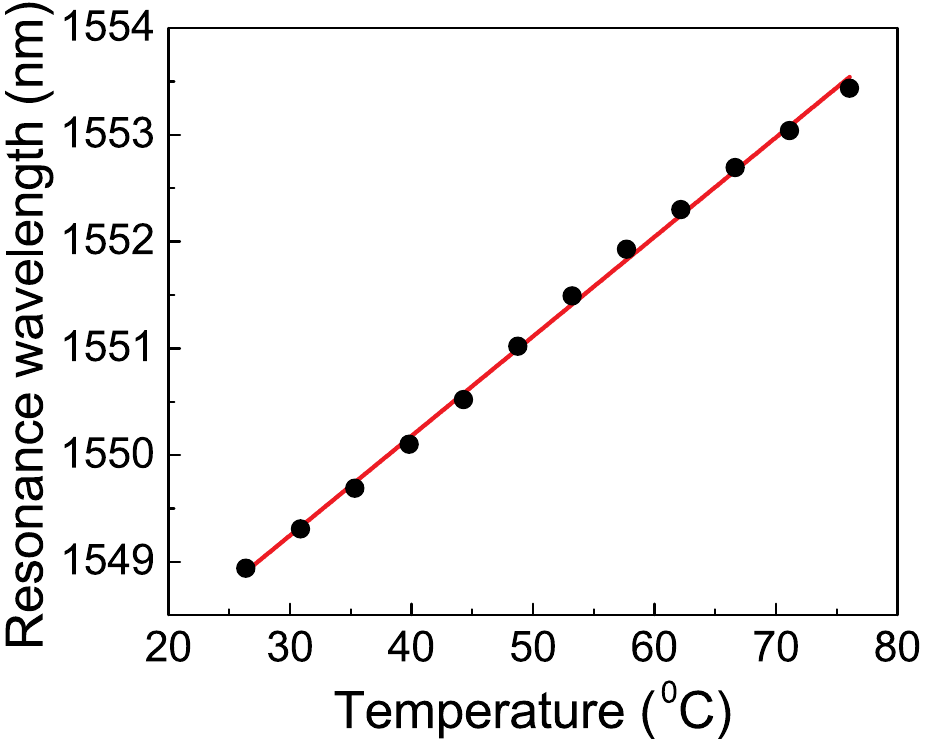}
 \caption{\label{fig3}\textbf{Resonance wavelength versus temperature of the sample.} The red line represents a line fit of the experimental data. The error for temperature is smaller than the datapoint size.}
\end{figure}

For the measurement of the thermo-optical coefficient the resonance wavelength was measured for several temperatures of the sample ranging from 26.4 to 76 $\rm{^oC}$. Spectra for all temperatures are presented in Figure \ref{fig2}. The resonance experiences a redshift which signifies that the material has a positive thermo-optical response. All spectra have a Fano-like line shape. There was no systematic change of the Q-factor which signifies that within this temperature range the mode-profile of the cavity does not change. In total the resonance redshifts by about 4.5 nm when the temperature rises by approximately 49.7 $\rm{^oC}$. The dependence of resonance wavelength on the temperature of the sample is presented in Figure \ref{fig3}. The resonance wavelength changes linearly with increasing temperature, as shown by the fit. 
 
When a sample is heated, there is always an extra resonance shift as a result of the evaporation of the water film from the surface of the sample \cite{oxydation1} in addition to the redshift due the increase in temperature. The exact magnitude of this extra shift depends on details of the surface condition. In laser heating experiments with similar temperature changes and resonance shifts we have observed this extra shift to be about 0.6 nm or less. As a worst case estimate we add an additional error term to the error of the linear fit which corresponds to the extra change of the resonance wavelength by 0.6 nm when temperature is increased by 49.7 $\rm{^oC}$. The resulting tuning slope is $d\lambda/dT=9\pm1\cdot 10^{-2}\, \rm{nm/\,K}$.

Using perturbation theory and scaling of Maxwell equations one can get a precise value of linear thermo-optic coefficient $dn/dT$ of the refractive index of Ga$_{0.51}$In$_{0.49}$P. To first order, the following equation for relative resonance wavelength shift can be applied \cite{joannopoulos}:
\begin{equation}\label{eq:1}
\frac{\Delta\lambda}{\lambda}=\frac{\Delta\lambda_n}{\lambda}+\frac{\Delta\lambda_a}{\lambda}
\end{equation}
Where $\Delta\lambda_n/\lambda$ and $\Delta\lambda_a/\lambda$ are relative resonance wavelength changes caused by two processes: refractive index increase and sample expansion due to heating. The first term is a direct result of the first order perturbation theory \cite{joannopoulos}, while the second term comes as a consequence of scaling nature of Maxwell equations \cite{joannopoulos}:
\begin{equation}\label{eq:2}
\frac{\Delta\lambda_n}{\lambda}=\frac{\Delta n}{n}\cdot\frac{\int\limits_{mebrane}^{}\varepsilon|\mathbf{E}(\mathbf{r})|^2d\mathbf{r}}{\int\limits_{all}^{}\varepsilon|\mathbf{E}(\mathbf{r})|^2d\mathbf{r}}=\frac{1}{n}\frac{dn}{dT}\Delta T\mathcal{E}_m
\end{equation}
\begin{equation}\label{eq:3}
\frac{\Delta\lambda_a}{\lambda}=\frac{\Delta a}{a}=\alpha_T\Delta T
\end{equation}
Here $\Delta n$ is the refractive index change due to temperature increase $\Delta T$, $\varepsilon$ and $n$ are the dielectric constant and refractive index of the membrane material and $\mathbf{E}(\mathbf{r})$ is the electric field of the cavity mode. $\mathcal{E}_m$ is the fraction of the electric-field energy inside the membrane. In Eq. \ref{eq:2} we took into account that the change in the refractive index of ambient nitrogen is negligible \cite{dndt_nitrogen} in comparison to the change of the refractive index of the semiconductor. According to our 3D FDTD calculations the fraction of electric-field energy in the membrane is 0.88. In Eq. \ref{eq:3} $\Delta a/a$ is the relative change of the photonic crystal lattice and $\alpha_T$ is the thermal expansion coefficient. Finally, the $dn/dT$ value can be obtained from:
\begin{equation}\label{eq:4}
\frac{dn}{dT}=\frac{n}{\lambda\mathcal{E}_m}(\frac{d\lambda}{dT}-\alpha_T\lambda)
\end{equation}

The thermal expansion coefficient \cite{Kudman1972} $\alpha_T$ for Ga$_{0.51}$In$_{0.49}$P is equal to $5.4\pm0.3\cdot10^{-6}\, \rm{K^{-1}}$. The experimentally measured value of $dn/dT$ is then equal to $dn/dT=2.0\pm0.3\cdot10^{-4}\,\rm{K^{-1}}$. We note that in case of GaInP linear expansion gives a noticeable contribution to the tuning slope of the cavity resonance. Without taking into account that effect the value of $dn/dT$ would be about 10\% larger.

In Ref. \citenum{Sokolov2015} we used complex modeling to estimate $dn/dT$ for a locally laser heated membrane, where the temperature was calculated from absolute power, thermal conductivity and absorptivity, thereby introducing many uncertainties. In addition, due to the local heating the membrane in that experiment was thermally stressed. The present experiment employs a direct temperature measurement with much less uncertain parameter performed on an unstressed membrane.     

\section{Conclusion}
In conclusion, we investigated the thermo-optical effect of the refractive index for  Ga$_{0.51}$In$_{0.49}$P. Our measurement took place for a freely expanding membrane and we took into account the effect of thermal expansion of the material. We found no significant Q-factor change during our measurement which guarantees that the working wavelength of photonic devices based on nanocavities made of Ga$_{0.51}$In$_{0.49}$P can be safely biased with temperature within the range of about 5 nm. This work enables precise thermal tuning of GaInP-based photonic devices, making GaInP one of the lowest thermal conductivity semiconductors used in photonics.\\

\textbf{Funding.} European Research Council project (ERC) (279248), Nederlandse Organisatie voor Wetenschappelijk.\\

\textbf{Acknowledgement.} The authors would like to thank Sanli Faez, Emre Y\"{u}ce and Willem Vos for helpful discussions and advises and Cornelis Harteveld for technical support.
\bibliographystyle{naturemag}

\end{document}